\documentclass[english,prl,twocolumn,showpacs,amsmath,amssymb,superscriptaddress,floatfix]{revtex4}
\usepackage[T1]{fontenc}
\usepackage{graphicx}

\usepackage{amsmath}
\usepackage{graphicx,epsfig}
\usepackage{amssymb}
\usepackage[dvips]{color}
\usepackage[T1]{fontenc}
\usepackage[latin9]{inputenc}
\usepackage{esint}
\usepackage{setspace}
\usepackage{bm}
\usepackage{pifont}

\makeatletter

\providecommand{\boldsymbol}[1]{\mbox{\boldmath $#1$}}

\usepackage{bm}
\usepackage{pifont}
\usepackage{babel}
\makeatother

\newcommand{\comment}[1]{}

\begin{document}

\preprint{}

\title{Tunable Kondo physics in a carbon nanotube double quantum dot}

\author{S. J. Chorley}
\affiliation{Cavendish Laboratory, University of Cambridge, Cambridge CB3 0HE, United Kingdom}

\author{M. R. Galpin}
\affiliation{Department of Chemistry, Physical and Theoretical Chemistry Laboratory, University of Oxford, Oxford OX1 3QZ, United Kingdom}

\author{F. W. Jayatilaka}
\affiliation{Department of Chemistry, Physical and Theoretical Chemistry Laboratory, University of Oxford, Oxford OX1 3QZ, United Kingdom}

\author{C. G. Smith}
\affiliation{Cavendish Laboratory, University of Cambridge, Cambridge CB3 0HE, United Kingdom}

\author{D. E. Logan}
\affiliation{Department of Chemistry, Physical and Theoretical Chemistry Laboratory, University of Oxford, Oxford OX1 3QZ, United Kingdom}

\author{M. R. Buitelaar}
\affiliation{Cavendish Laboratory, University of Cambridge, Cambridge CB3 0HE, United Kingdom}

\date{\today}

\begin{abstract}
We investigate a tunable two-impurity Kondo system in a strongly correlated carbon
nanotube double quantum dot, accessing the full range of charge regimes. In the regime
where both dots contain an unpaired electron, the system approaches the two-impurity Kondo
model. At zero magnetic field the interdot coupling disrupts the Kondo physics and a local
singlet state arises, but we are able to tune the crossover to a Kondo screened phase by
application of a magnetic field. All results show good agreement with a numerical
renormalization group study of the device.
\end{abstract}


\maketitle


%
A quantum dot coupled to two leads can be considered as an
experimental realization of the single-impurity Anderson model
\cite{Anderson}. When the dot contains a single unpaired
electron, the Anderson model accurately describes how, below a
characteristic temperature $T_K$, correlated electron tunnelling
between the quantum dot and the leads results in coherent
screening of the electron spin \cite{Goldhaber,Cronenwett,Nygard}.
The combined system of electrons on the quantum dot and leads
forms a spin singlet, a phenomenon known as the Kondo effect \cite{Hewson}.
Likewise, two tunnel-coupled quantum dots should amount to an experimental
realization of the two-impurity Anderson model~\cite{Alexander}. Here the
physics is much richer, particularly in the regime where each dot
contains an unpaired electron. In this case, a competition now arises
between the tendency of the conduction electrons on the leads
to screen the spins on the quantum dots, and the antiferromagnetic
exchange coupling $J$ between the two localized spins. The former
favors formation of a Kondo singlet between each lead and the
dot to which it is coupled, while the latter favors a local singlet
state. The resulting groundstate of the system depends sensitively
on the relative strength of the interactions, an understanding of
which is important and believed to underlie the electronic
properties of a wide range of strongly correlated materials,
including spin glasses and heavy fermion compounds \cite{Hewson}.

The essence of this competition is captured by the two-impurity
Kondo model \cite{Jaya,Jones1,Afflec92,Afflec95}, which describes
the low-energy physics of the two-impurity Anderson model in the
absence of charge transfer between the leads, and famously contains
a quantum phase transition at the boundary of the local and Kondo singlet
phases where $J \sim T_K$. While this has attracted considerable
experimental attention \cite{Jeong,Craig,Bork}, observation of the
transition has remained elusive. This is perhaps unsurprising as
charge transfer between the leads, absent in the two-impurity Kondo model,
is necessarily present in experiment if a conductance is measured. As
is well known theoretically, this transforms the quantum phase transition
into a crossover, such that the ground state of the system is always a Fermi
liquid \cite{Afflec95, Jayatilaka}; although remnants of the transition are
evident in a strong enhancement of the zero-bias conductance in the vicinity
of $J\sim T_{K}$ \cite{Georges,saka:2000,sela:2009}.

An understanding of the transport properties of a realistic
two-impurity system, such as a double quantum dot (DQD), thus requires
that charge transfer between the leads is taken into account. This
is achieved in the present work, where we present a study of a tunable
carbon nanotube DQD in the strongly correlated regime, and use a
numerical renormalization group (NRG) study of the two-impurity
Anderson model to describe the device. We show that in the charge
regime where both dots have an unpaired electron, the ground state
of the device is a local singlet phase with suppressed Kondo correlations.
The ability to tune the exchange coupling and the Kondo scales allows
one, in principle, to crossover from the local singlet to the Kondo
screened phase. In our device the onset of charge fluctuations prevents
the crossover being seen cleanly at zero magnetic field. We have, however,
been able to observe it at finite magnetic field, consistent with recent
theoretical predictions \cite{Jayatilaka}.

The device we consider is a single-walled carbon nanotube on a
degenerately doped Si/SiO$_2$ substrate contacted by Au contacts,
see Fig.~1(a)~\cite{epaps}. A central gate is used to introduce a tunable
tunnel barrier, separating the nanotube into two quantum dots, which can be
individually addressed by two additional side gates. The stability
diagram of the device is shown in Fig. 1(b). The effective electron
number of each charge regime is indicated by the ordered pairs
$(n,m)$. The large-small-large alternation in the stability diagram
reflects the two-fold spin degeneracy of the device, and allows us to
establish unambiguously the parity of the electron number (even or
odd). The orbital degeneracy of the nanotubes is broken, most likely
as a result of $K - K'$ mixing \cite{Liang, Buitelaar1}.

From an analysis of the stability diagram we are able to extract the
on-site charging energies $U \sim 2.5$ meV for both dots, the
electrostatic coupling energy $U' \sim 0.6$ meV and the interdot
tunnel coupling $t \sim 0.4$ meV. The widths of the Coulomb blockade
peaks in the stability diagram allow an estimate of the coupling $\Gamma$
between the dots and the leads. In the non-interacting limit the full-width
at half maximum is equal to $2\Gamma$, but the peaks are broadened
by a further factor of $\sim 2$ by many-body scattering processes~\cite{Block}.
Bearing this in mind, we obtain $\Gamma_R \sim 0.23$ meV and
$\Gamma_L \sim 0.12$ meV for the right and left dot respectively.
For these coupling strengths $U/\Gamma_{\nu} \sim 10$,  so we might
expect co-tunneling processes, and thus the Kondo effect, to be experimentally
observable. To investigate the presence of Kondo correlations, we measured the
differential conductance in all charge regimes, see Fig.~1(c) and 1(d). When both
dots contain an even number of electrons, the Kondo effect is inoperative and
conductance is suppressed and featureless around source-drain bias $V_{sd}=0$.
On the other hand, when \textit{one} of the quantum dots contains an odd number
of electrons, we observe a pronounced zero-bias conductance peak. The
evolution of the differential conductance as electrons are added to
the right dot, keeping the effective electron number of the left dot
fixed at $N=2$, is shown in the top-most panel of Fig.~1(d). The
characteristic appearance of a zero-bias peak when the electron
number is odd is a clear indication of Kondo physics in the (0,1),
(1,0), (1,2) and (2,1) charge regimes \cite{Goldhaber,Cronenwett,Nygard},
the spin of the singly-occupied dot being effectively Kondo screened by
the lead to which it is coupled.

The behavior, however, is markedly different in the center of the (1,1)
charge regime where the electron number is odd for \textit{both} dots. As
shown in Fig.~1(c), the zero-bias conductance is suppressed and a double peak
structure arises at finite bias \cite{Jeong,Craig}. The behavior we observe
is characteristic of a strongly correlated DQD when $J \gg T_K^L, T_K^R$ where
$T_K^L, T_K^R$ denote the Kondo scales for the left and right dots respectively.
These energy scales are readily estimated using the parameters obtained
above. In the center of the (1,1) charge regime the exchange  coupling
$J \sim 4t^2/(U-U')$, which yields $J\sim 0.34$ meV. The Kondo scales can be
estimated roughly using the Haldane expression \cite{Haldane}
\begin{equation}
\label{Haldane}
T_K^{\nu} \sim \sqrt{\Gamma_{\nu} U} \textrm{exp}[\pi \epsilon_{\nu}(\epsilon_{\nu}+U)/2 \Gamma_{\nu} U]
\end{equation}
where $\epsilon_{\nu}$ ($\nu  = L,R$) is the level energy of dot $\nu$ relative to the zero-bias Fermi level of the leads. This yields $T_K^R\sim 10^{-2}$ meV and $T_K^L\sim 10^{-4}$ meV in the middle of the (1,1) charge regime, such that $J \gg T_K^L, T_K^R$.

\begin{figure}
\includegraphics[width=85mm]{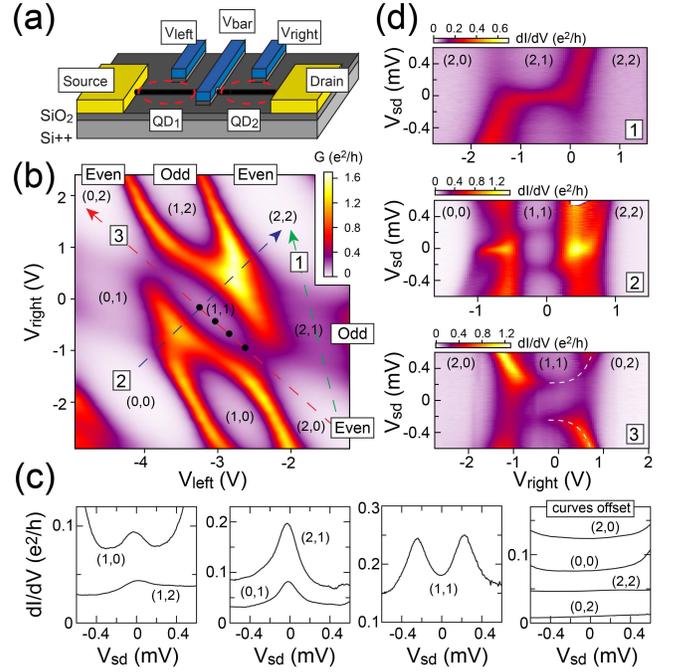}
\caption{\label{Fig1}(color online) \textbf{(a)} Schematic of the carbon nanotube DQD device.\textbf{(b)} Stability
diagram of the DQD measured at $T \sim 60$ mK. The Si backgate voltage $V_{bg} = -1.7$ V. The barrier gate voltage $V_{bar} = 0$ mV. \textbf{(c)} Differential conductance in the centers of the various charge regimes of panel (b) as indicated by $(n,m)$. The curves in the rightmost panel are offset in steps of 0.03 $e^2/h$. \textbf{(d)} Differential conductance along the lines indicated in panel (a).}
\end{figure}

\begin{figure}
\includegraphics[width=83mm]{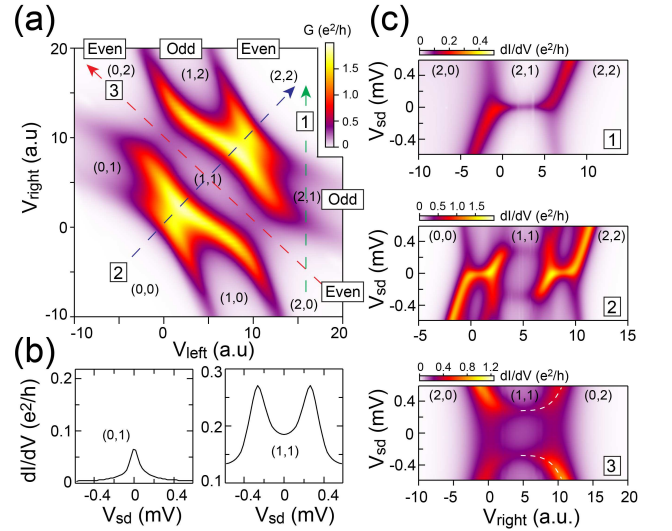}
\caption{\label{Fig2}(color online) \textbf{(a)} NRG calculation of the stability diagram. \textbf{(b)} Differential conductance in the centers of the various charge regimes of panel (a) as indicated by $(n,m)$. \textbf{(c)} Differential conductance along the lines indicated in panel (a).}
\end{figure}

The above interpretation of the measurements is strengthened by an NRG
study of our double quantum dot. The calculated stability diagram using the
experimental parameters, see Fig.~2(a), reproduces all key features of the
experiment and allows for a comparison between experiment and theory (for
further details see supplementary material \cite{epaps}).
Finite-bias conductances, NRG calculation of which is inevitably approximate
(see \cite{epaps}), are also in good agreement. In particular the double peak
structure in the (1,1) charge regime is reproduced in the calculations, see Fig.~2(b).
It can be understood physically as a non-equilibrium Kondo effect \cite{Paaske}. While
the formation of a local singlet state suppresses Kondo correlations at zero-bias,
interlead spin-flip tunneling becomes possible and Kondo correlations are partially
restored when $V_{sd}$ is comparable to the exchange energy $J$ separating the
atomic-limit singlet ground state from its triplet excited states. As demonstrated
below, a unique feature of double quantum dots is that $T_K^L, T_K^R$ and $J$ are
all tunable by varying the dot level energies.

We discuss first the behavior along diagonal 2 in Fig.~1(b), for which
the differential conductance is shown in Fig.~1(d), middle
panel. Along this diagonal, the dot level energies $\epsilon_{L,R}$
are decreased while their difference (or detuning) $\epsilon_L -\epsilon_R$
remains zero. This allows us to tune the Kondo scales of the two quantum dots,
as these strongly depend on the dot level energies, see  Eq.\eqref{Haldane}.
In the center of the (1,1) charge regime, the Kondo scales for both dots are at a
minimum. Moving away from the center, the Kondo scales thus increase. However,
before we can access the Kondo screened phase we reach the edge of the (1,1) charge
regime, moving into a mixed-valence regime. Here charge fluctuations are significant, and
a two-impurity Kondo model description is no longer valid. The inability to access the
Kondo screened phase this way can be understood given that for our device $J \sim \Gamma_{L,R}$,
so to observe the crossover requires $T_K^{L,R} \sim \Gamma_{L,R}$, which occurs only when
$|\epsilon_{L,R}| \lesssim \Gamma_{L,R}$, i.e. in the mixed-valence regime. A lower $J$ is
therefore required in order to observe the crossover cleanly in this way. While we were able to
control the tunnel coupling $t$ experimentally (see ~\cite{epaps}), sufficiently small values of
it were not reached here.

\begin{figure}
\includegraphics[width=85mm]{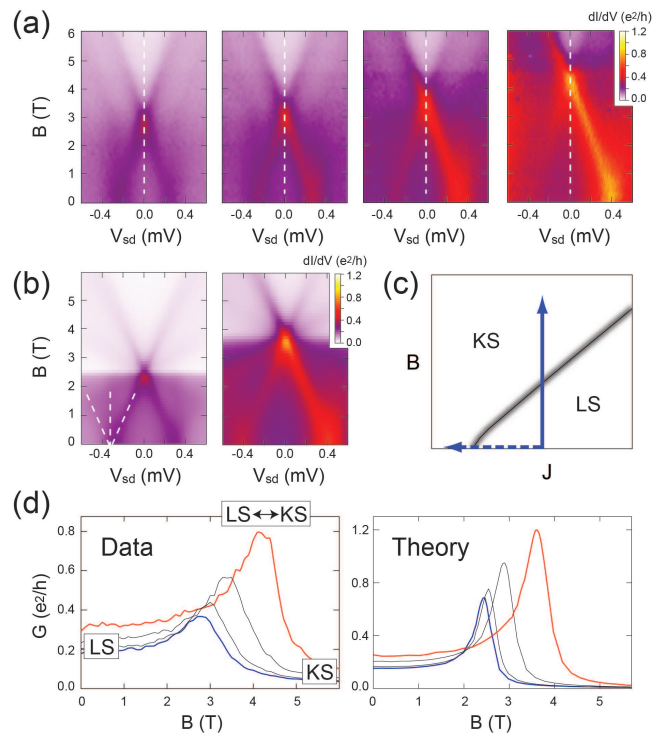}
\caption{\label{Fig3} (color online) \textbf{(a)} Magnetic field dependence of the differential conductance
in the (1,1) charge regime for various values of the detuning as indicated in Fig.~1(b) (black dots along line 3), with the left plot for the center of the (1,1) regime.
\textbf{(b)} NRG results corresponding to the left and right panels in (a). Dashed lines show the threefold splitting of one zero-field peak.
\textbf{(c)} Schematic of crossover between LS and KS phases, see text, either by varying $B$ (solid arrowed line) or $J$ (dashed arrowed line).
\textbf{(d)} Zero-bias conductance \emph{vs} $B$ for different detuning, along dashed white lines in panel (a); both measured (left) and calculated (right). Individual curves from left to right correspond to (a) from left to right.}
\end{figure}

We now focus on the behavior along diagonal 3 in Fig.~1(b) for which
the differential conductance is shown in Fig.~1(d), bottom panel.
Along this diagonal the detuning ($\epsilon_L - \epsilon_R$) is
increased while $\epsilon_L + \epsilon_R$ is constant. As indicated
by the dashed white line in Fig.~1(d) the peaks observed in the
differential conductance move further apart as the magnitude of the
detuning is increased and the conductance becomes highly asymmetric
in bias. This increase in peak splitting is due to  $J$ becoming
larger with positive detuning as the (1,1) singlet state
becomes closer in energy to the (0,2) singlet \cite{J}. The ability to
tune the exchange energy by varying the detuning is essential in spin-based
quantum information processing schemes using quantum dots \cite{Chorley}.
Importantly, the present data shows that this tunability can also be used
as a probe of Kondo physics.

The consequences of varying the exchange $J$ are investigated further in
Fig.~3 by application of a magnetic field ($B$) perpendicular to the nanotube axis,
for various values of the detuning along diagonal 3 in the $(1,1)$ regime.
At finite $B$ the observed zero-field peaks split into three components, Fig.~3(a)
(of which the innermost peaks are most easily resolved, see Figs.~3(a,b)), consistent
with a local singlet ground state and triplet excited state, separated by $\sim J$ at $B=0$.
As $B$ is increased, the energy difference ($\sim$$J -g\mu_{B}B$) between the singlet ($S$)
and the lowest-energy triplet state, $T_-$, decreases. At a field $B_c \sim J/g\mu_B$ these
states are near-degenerate. A strong zero-bias conductance peak is then observed, due to
Kondo screening of the  $S$-$T_-$ pseudospin system \cite{PustKikoin,PustGlazman,Florens},
here  by two channels. [In our tunnel-coupled double QD both the $S$ and $T_{-}$ states arise
from the $(1,1)$ charge configuration, in contrast to $S$-$T_{-}$ crossings observed \cite{Nygard,Paaske,Zumbuhl,Roch} in single QDs with two `active' levels, where the relevant
singlet is a configuration with the lower dot level doubly occupied \cite{faraway},
and consequent Kondo screening arises by a single effective channel \cite{PustKikoin}.]

From the perspective of the two-impurity Anderson model, and its experimental realization in
our device, this conductance peak is the signature of the finite-field crossover from a local
singlet (LS) phase for $B\lesssim B_{c}$, to a polarized Kondo screened (KS) phase for
$B \gtrsim B_{c}$; and amounts to a finite-$B$ continuation of the zero-field conductance peak
at $J \sim T_{K}$~\cite{Jayatilaka, Georges,saka:2000,sela:2009}.
For a L-R symmetric two-impurity Kondo model, it was recently shown \cite{Jayatilaka}
that the well known zero-field quantum phase transition between LS and KS phases,
occurring at $J=J_{c} \sim T_{K}$, extends into the $(B, J)$-plane (Fig. 3(c)).
The transition could thus be driven either by tuning $J$ through $J_{c}$ at zero-field,
\emph{or} by tuning $B$ through a critical $B_{c}$ at fixed $J$. With interlead charge
transfer and L-R asymmetry, both of which occur in our device, the transition line is
of course broadened into a line of crossovers (Fig. 3(c)) with associated conductance peaks.
For $J \gg J_{c}$ in particular, where the system at zero-field is deep in the LS phase,
$B_{c} \sim J/g\mu_{B}$ (and the Kondo screened phase for $B>B_{c}$ is naturally spin
polarized, asymptotically approaching the free $T_{-}$ state). This behavior is observed
clearly in Fig.~3(d). In the center of the (1,1) regime, $J\sim 0.34$ meV (as estimated
above) yields $B_{c}\sim J/g\mu_{B}  = 2.9$ T, in very good
agreement with the measured conductance peak in Fig.~3(d)~\cite{epaps}.
As detuning increases, $J$ becomes
larger as noted above, and $B_c$ is correspondingly seen to increase. The evolution of conductance
as a function of magnetic field is in good agreement with the NRG calculations as shown in Fig.~3(b,d).
The calculations also confirm (not shown) that the asymmetry in $V_{\mathrm{sd}}$ observed (Fig.~3(a))
in the differential conductance with increasing detuning, results from the asymmetry
$\Gamma_L \neq \Gamma_R$. For symmetric coupling strengths, or precisely at the center of the (1,1)
charge regime where the system is electron-hole symmetric [e.g. the leftmost plot in Fig.~3(a)],
no asymmetry is observed in the differential conductance.

\begin{figure}
\includegraphics[width=85mm]{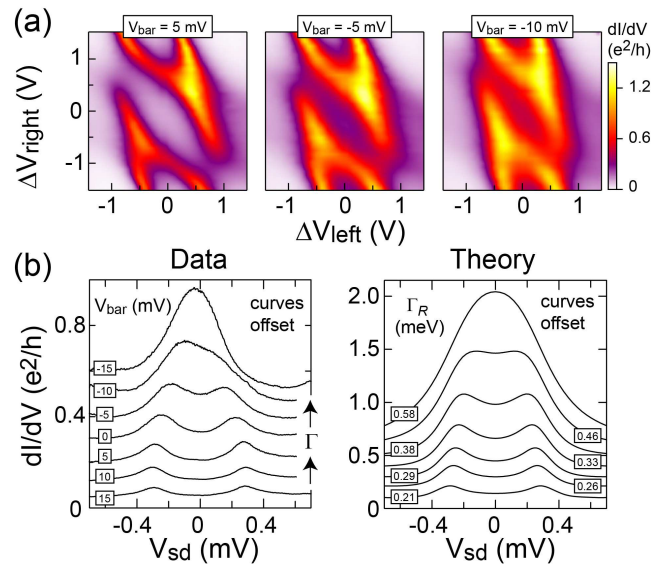}
\caption{\label{Fig4}(color online) \textbf{(a)} Observed stability diagrams of the DQD for different dot-lead coupling strengths as varied by $V_{bar}$.
\textbf{(b)} Left: differential conductance in the center of the (1,1) charge regime for different
dot-lead couplings. The curves are offset in steps of 0.05 $e^2/h$. Right: NRG calculations, with
the ratio $\Gamma_R / \Gamma_L = 2$ fixed and $\Gamma_{R}$ as indicated (curves again offset in steps of 0.05 $e^2/h$).}
\end{figure}

Finally, we show that by varying $V_{bar}$ (see Fig.~1(a)) we can directly tune the coupling strengths between the quantum dots and their leads \cite{coupling}. This is illustrated in Fig.~4(a) which shows that for decreasing $V_{bar}$, the conductance at the center of the (1,1) charge regime strongly increases. At the same time, the double peak structure, clearly observed for higher values of $V_{bar}$, gradually merges into one broad peak.
This behavior can be understood as an increase in the coupling strengths $\Gamma_{L,R}$, as illustrated by NRG calculations, see Fig.~4(b), and consistent with the increase in widths of the Coulomb blockade peaks in the stability diagrams. On increasing $\Gamma_{L,R}$, and hence decreasing $U/\Gamma_{L,R}$, the device becomes less strongly correlated and charge fluctuations consequently more significant, thereby eroding and ultimately destroying the double peak structure that is characteristic of the strongly correlated regime of the DQD.

In conclusion, we have investigated a tunable two-impurity Kondo system in a strongly correlated carbon
nanotube DQD, in which the full range of charge regimes is accessible, and which amounts to a clear
experimental realization of a two-impurity Anderson model.
In the (1,1) regime, we have shown that the exchange and Kondo energy scales can be varied by tuning
the dot energy levels or by varying the dot-lead tunnel couplings, providing the possibility of observing
the crossover between local singlet and Kondo screened phases. While charge fluctuations in this device
prevented observation of the crossover at zero field, we were able to observe it at finite field, indicated by enhanced zero-bias conductance. The work is readily extended to carbon nanotube DQDs coupled to superconducting \cite{Buitelaar2} or ferromagnetic \cite{Hauptmann} leads. This allows experimental access to
rich phase behavior controlled by the interplay between Kondo, exchange, and superconducting correlations, and further highlights the potential of carbon nanotube quantum dots to investigate correlated electron physics.

\emph{Acknowledgments.--} We thank David Cobden and Jiang Wei for the carbon nanotube growth.
M.R.B. is supported by the Royal Society, and D.E.L. acknowledges support from EPSRC through
EP/I032487/1.




\onecolumngrid
\newpage

\setcounter{page}{1} \thispagestyle{empty}


\begin{center}
\textbf{{\large Tunable Kondo physics in a carbon nanotube double quantum dot:\\
Supplementary Material}}\\
\bigskip
S. J. Chorley,$^{1}$ M. R. Galpin,$^{2}$ F. W. Jayatilaka,$^{2}$ C.
G. Smith,$^{1}$ D. E. Logan$^{2}$ and M. R. Buitelaar$^{1}$

\textit{$^{1}$Cavendish Laboratory, University of
Cambridge, Cambridge CB3 0HE, United Kingdom\\
$^2$Department of Chemistry, Physical and Theoretical Chemistry
Laboratory,\\ Oxford University, South Parks Road, Oxford OX1 3QZ,
United Kingdom}

\end{center}

\setcounter{figure}{0}
\setcounter{equation}{0}

\renewcommand{\figurename}{Figure S}



\providecommand{\boldsymbol}[1]{\mbox{\boldmath $#1$}}





%
%
%
%

The following appendices comprise the supplementary material to
Ref.~\onlinecite{Spaper}. We provide the following: (i) details of
device fabrication; (ii) additional experimental data on the
dependence of the double quantum dot (DQD) conductance on the
applied magnetic field, temperature and interdot and dot-lead tunnel
coupling; and (iii) details of the model Hamiltonian used in the
numerical renormalization group (NRG) calculations, and how the
conductance is calculated from the NRG.


\section{Device fabrication}
Carbon nanotubes were synthesized by chemical vapor deposition using
a procedure similar to Ref.~\onlinecite{SKong}. To form the catalyst
particles, $\sim 0.5$  $\mu$g/cm$^{3}$ Fe(NO$_3$).9H$_2$O
(Sigma-Aldrich) was sonicated in isopropanol. The bare Si/SiO$_2$
substrate (300 nm thermal oxide) was dipped into the solution
immediately after sonication and dried by air blowing. It was then
heated to 900 $^{\circ}$C in a 1" tube furnace under a hydrogen gas
flow of 400 sccm, during which the iron nitrate was reduced to iron
particles. Once at this temperature, 500 sccm of methane (containing
natural isotope ratios) was added to the flow for 8 minutes. After
the methane flow was stopped, the substrate was cooled under the
same 400 sccm hydrogen flow. Alignment marks were defined by
electron-beam lithography, and scanning electron microscopy and
atomic force microscopy were used to locate the nanotubes in
relation to these. Further steps of electron-beam lithography and
evaporation contact the nanotubes with 20 nm thick, 300 nm wide gold
ohmic contacts, separated by 700 nm, which also form the outer
barriers of the quantum dots. The final electron-beam lithography
layer writes the 100 nm wide and 100 nm spaced gates, which are made
from two 1.2 nm layers of aluminum, oxidized in air and capped with
a further 20 nm of titanium and 6 nm of gold. The source lead is
grounded by the virtual earth of a $\times 10^7$ current to voltage
preamplifier, and the drain has a small variable dc bias applied.
The chip is mounted on the cold finger of a 60 mK dilution
refrigerator and connected to the measurement apparatus with
filtered low-frequency lines.

\begin{figure}
\includegraphics[width=155mm]{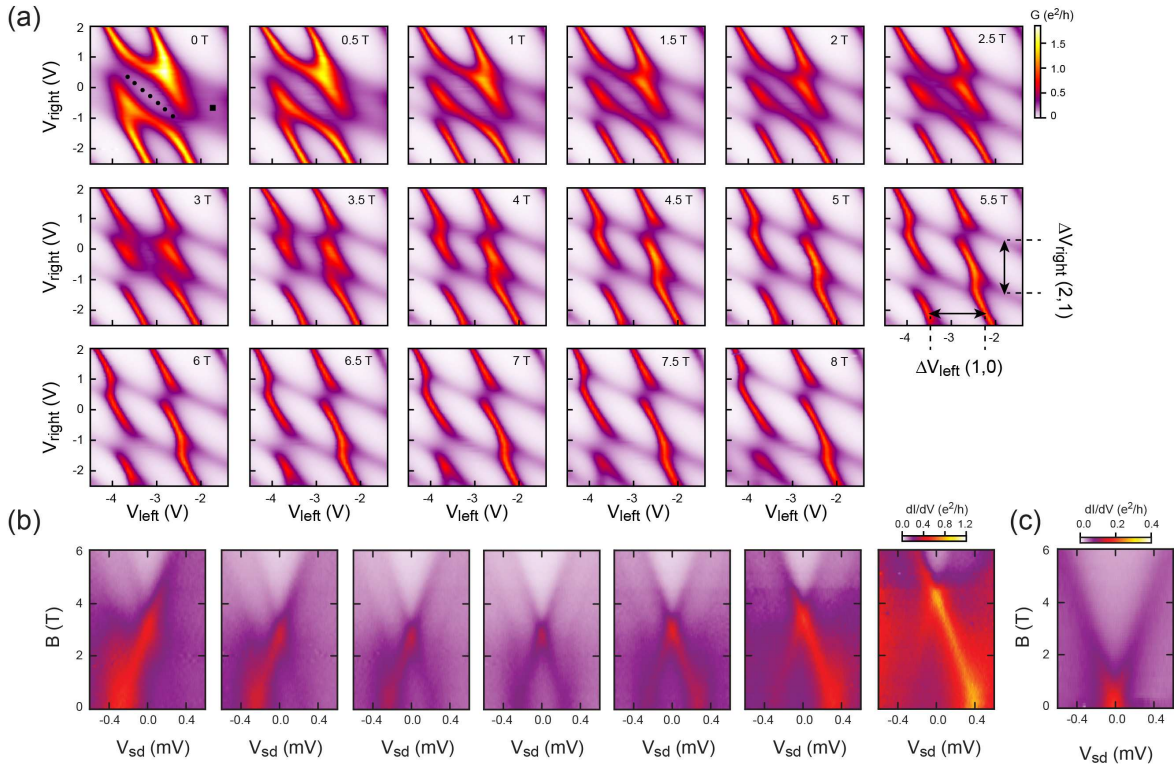}
\caption{\label{Fig1}\textbf{(a)} Double quantum dot stability
diagrams measured at $T\sim 60$ mK. The magnetic field is increased from 0 in
steps of 0.5 T as indicated.
\textbf{(b)} Differential conductance as a function of
magnetic field at positions along the diagonal of the (1,1) charge
region as indicated by the closed circles ($\bullet$) in the $B=0$
diagram of panel (a).
\textbf{(c)} Differential conductance as a
function of magnetic field in the center of the (2,1) charge region
as indicated by the closed square ($\blacksquare$) in the $B=0$
diagram of panel (a).}
\end{figure}

\section{Magnetic field and temperature behavior}

Figure S1(a) shows the evolution of the double quantum dot stability
diagram with magnetic field in steps of 0.5 T. We first focus on the
widths of the hexagons (1,0) and (2,1). In the atomic limit (no
coupling between the dots and their leads) the width of the hexagon
increases proportional to $g \mu_B B$ for a magnetic field applied
perpendicularly to the nanotube axis, where we assume $g=2$ for nanotubes.
In practice our device has a relatively strong coupling
between the dot and leads, and a linear relation between the
width of the hexagons and $B$ is a good approximation only in the
limit of large magnetic field. Bearing this in mind, we obtain the
coupling strengths (or lever arms) between the gate electrodes and
the quantum dots, and extract charging energies $U \sim 2.5$ meV
for both quantum dots and an electrostatic coupling energy $U'\sim 0.6$
meV.

The differential conductance as a function of magnetic field is
shown in Figs.~S1(b) and (c) - which are an extension of Fig.~3 in
the main text - for different positions in the stability diagram.
Fig.~S1(b) shows the differential conductance as a function of
magnetic field for seven positions in the (1,1) charge region as
indicated by the circles in the $B=0$ diagram in panel (a).
Fig.~S1(c) shows the differential conductance as a function of
magnetic field in the center of the (2,1) charge region as indicated
by the square in the $B=0$ diagram in panel (a). The maximum
conductance in the center of the (1,1) charge region is observed at
a magnetic field $B \sim 2.8$ T as seen both in Fig.~S1(a) and (b).
The maximum conductance in the center of the (2,1) charge region is
observed at $B = 0$ T, as expected, with an approximately linear
splitting of the Kondo resonance as the magnetic field is increased.

\begin{figure}
\includegraphics[width=80mm]{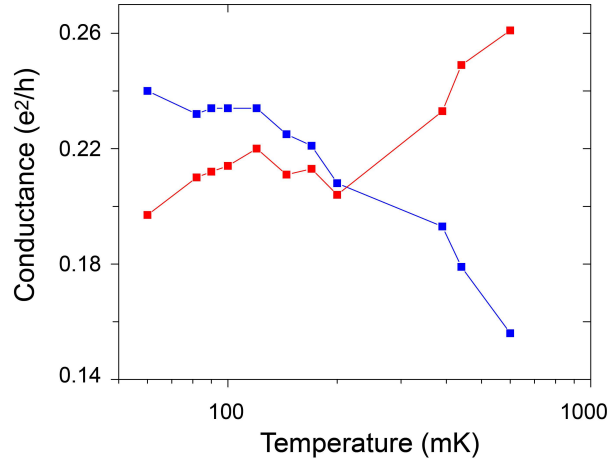}
\caption{\label{Fig2}Measured temperature dependence of the conductance in the
center of the (1,1) charge region (red points) and (2,1) charge region (blue points).}
\end{figure}

The temperature dependences of the zero-bias conductance in the centers of the (1,1)
and (2,1) charge regions are shown in Fig. S2. They can readily be explained with
reference to Fig. 1(c) of the main text, noting that the dominant effect of increasing
temperature is to broaden the low-energy conductance peaks.
In the center of the (1,1) region (red points) there is a gradual increase of
zero-bias conductance with temperature, corresponding to a filling-in of the
conductance dip in the third panel of Fig. 1(c). In the (2,1) region (blue points)
by contrast, the broadening of the single conductance peak (2nd panel of Fig. 1(c))
with temperature results in a gradual decrease in the zero-bias conductance. NRG
results for similar parameters (not shown) display the same qualitative trends.

\begin{figure}
\includegraphics[width=150mm]{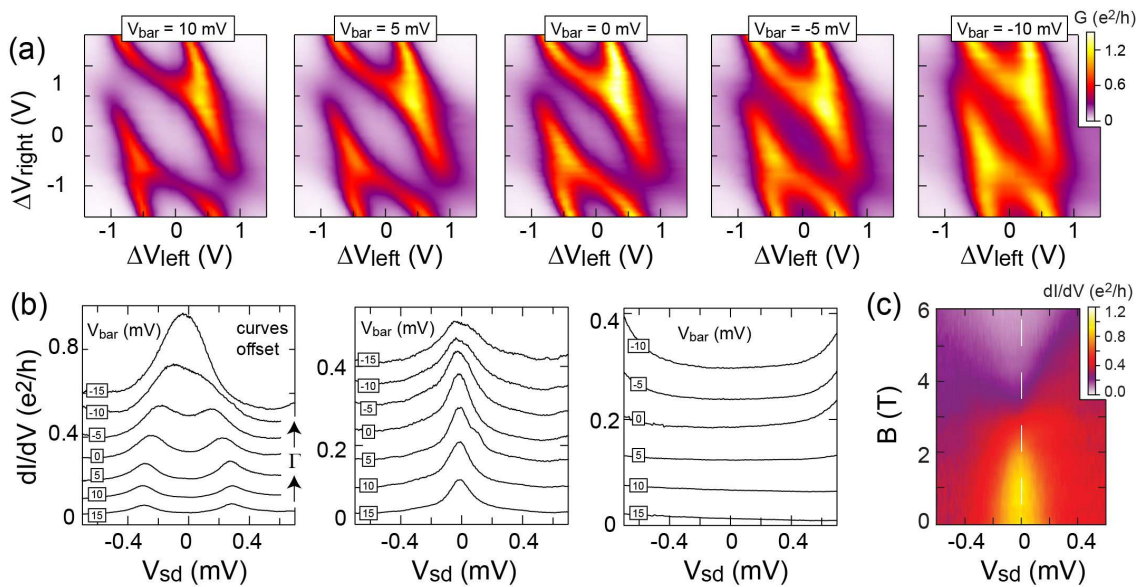}
\caption{\label{Fig3}\textbf{(a)} Stability diagrams of the double
quantum dot for five different barrier gate voltages $V_{bar}$.
\textbf{(b)} Differential conductance in the center of the (1,1)
charge region (left), the (2,1) charge region (middle) and (2,0)
charge region (right) for a series of barrier gate voltages as
indicated. Curves are offset in steps of 0.05 $e^2/h$ in each panel.
\textbf{(c)} Differential conductance as a function of magnetic
field in the center of the (1,1) charge region for $V_{bar} = -15$ mV.}
\end{figure}

\section{Varying dot-lead and dot-dot coupling strengths}

By varying the barrier gate voltage $V_{bar}$ we were able to tune
both the interdot coupling and the coupling between the dot and
leads. Which of these two effects dominates varies between different
electronic states, the microscopic details of which are not
presently understood. Nevertheless, the effects are highly
reproducible and can be used to investigate the competition between
exchange and Kondo correlations on the quantum dots.

Figure S3 shows the stability and differential conductance in the
center of the (1,1) charge region as $V_{bar}$ is varied - an
extension of Fig.~4 in the main text. In this case, the main effect
of varying $V_{bar}$ is an increase in the coupling to the leads.
This is deduced from the increase in width (FWHM) of the Coulomb
blockade peaks in the stability diagram as well as the increased
width of the Kondo resonance observed in the (2,1) charge region,
see middle panel of Fig.~S3(b). The double peaks in the center of
the (1,1) charge region broaden with decreasing $V_{bar}$ to form a
single broad peak centered around $V_{sd}=0$, interpreted as a
crossover to the weakly correlated regime where $U/\Gamma_{L,R}
\not\gg 1$, see main text.

A different dependence on $V_{bar}$ is observed for the neighboring
`effective' (1,1) charge region, see Fig.~S4. As expected, a double
peak structure is observed in the differential conductance. However,
in this case, the interdot coupling $t$ is changed, resulting in a
decrease of the exchange energy $J$ with increasing $V_{bar}$ and
correspondingly smaller peak splitting, see Fig.~S4(c). The change
in $t$ is also evident in the stability diagram, see Fig.~S4(b),
where it is observed as a decrease in the separation (and curvature)
of the triple points. The smallest peak splitting in the
differential conductance was observed for $V_{bar} = 15$ mV, see
Fig.~S4(c). No further decrease in $t$ was observed by increasing
$V_{bar}$ beyond this value (not shown).

\begin{figure}
\includegraphics[width=130mm]{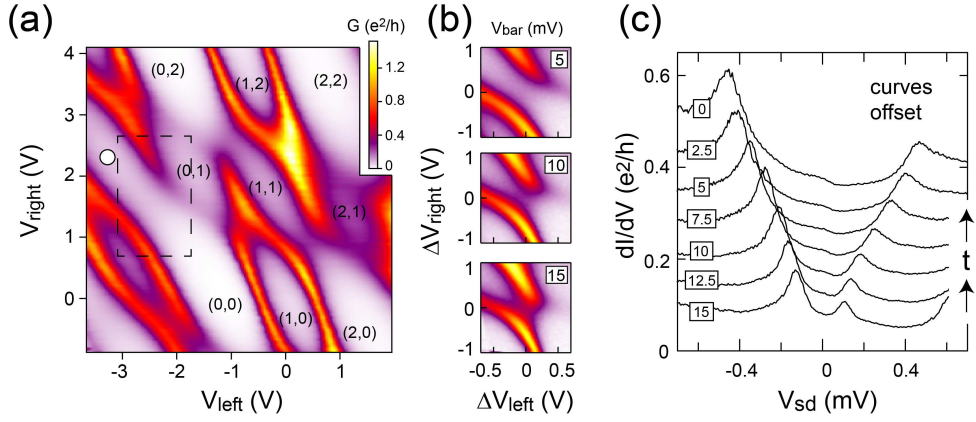}
\caption{\label{Fig4}\textbf{(a)} Stability diagram of the double quantum dot. Here
the Si backgate voltage is set to $V_{bg} = -1.8$ V and $V_{bar}=0$ mV. The circle
indicates the approximate center of an effective (1,1) charge region, different
from that discussed in the main text.
\textbf{(b)} Conductance around a triple point pair for the area indicated in
panel (a) by the dashed lines for three different $V_{bar}=5, 10$, and 15 mV.
\textbf{(c)} Differential conductance in the center of the leftmost (1,1) charge
region, i.e., position indicated by the circle in panel (a), for a series of $V_{bar}$. The
increase of the peak splitting correlates with an increase of the interdot coupling $t$.}
\end{figure}

\section{Model Hamiltonian}
The canonical model used to describe tunnel-coupled DQDs is the
two-impurity Anderson
model~\cite{Salex:64,Sgott:64,Syama:79,Sjaya:82}. We take this as
the basic model for our NRG calculations, and supplement it with an
\emph{inter}dot Coulomb repulsion/electrostatic coupling term
($\hat{H}_{U'}$), which inevitably exists, and which we find
necessary to obtain good agreement between theory and the
experimental stability diagram shown in Fig.~1(b).

The Hamiltonian, $\hat{H} = \hat{H}_{\mathrm{L}} + \hat{H}_{\mathrm{D}} + \hat{H}_{\mathrm{hyb}}$, consists
of terms for the isolated leads and dots, $\hat{H}_{\mathrm{L}}$
and $\hat{H}_{\mathrm{D}}$ respectively, together with a hybridization term, $\hat{H}_{\mathrm{hyb}}$, describing
electron tunneling between dot-$\nu$ ($\nu =L,R$) and lead-$\nu$.
The leads, for which
\begin{equation}
\hat{H}_{\mathrm{L}} = \sum_{\nu, \mathbf{k},\sigma} \epsilon_{\mathbf{k}}^{\phantom\dagger} c^{\dag}_{\nu \mathbf{k}\sigma}c^{\phantom{\dag}}_{\nu \mathbf{k}\sigma},
\end{equation}
consist of equivalent bands of non-interacting electrons with
single-particle energies $\epsilon_{\mathbf{k}}$ and spin $\sigma
=\uparrow,\downarrow$ (where $c^{\dag}_{\nu \mathbf{k}\sigma}$
creates a $\sigma$-spin electron in state $\mathbf{k}$ of lead
$\nu$). As usual for metallic
leads~\cite{Swils:75,Skris:80,Skris2:80,Sbull:08}, we take the leads
to be flat bands with half-bandwidth $D$. For the dots,
\begin{equation}
\begin{split}
\hat{H}_{\mathrm{D}} = &\sum_{\nu,\sigma}\epsilon_{\nu}\hat{n}_{\nu \sigma} ~+~ t\sum_{\sigma} \left({d}^{\dag}_{L\sigma}{d}^{\phantom{\dag}}_{R \sigma} + h.c. \right) \\
&+ ~U\sum_{\nu}  \hat{n}_{\nu \uparrow}\hat{n}_{\nu \downarrow} ~+~ U'\hat{n}_L \hat{n}_R~,
\end{split}
\end{equation}
where $\hat{n}_{\nu \sigma}= d^{\dag}_{\nu\sigma}d^{\phantom{\dag}}_{\nu \sigma}$ is the
number operator for $\sigma$-spin electrons on dot $\nu$, and
$\hat{n}_{\nu} = \hat{n}_{\nu \uparrow} + \hat{n}_{\nu \downarrow}$.
$\epsilon_{\nu}$ is the single-particle level energy for dot-$\nu$, $t$ the interdot tunnel coupling, $U$ the on-dot charging energy/Coulomb repulsion, and $U'$ the interdot Coulomb interaction. The hybridization term coupling the dots to the leads is given by
\begin{equation}
\hat{H}_{\mathrm{hyb}} = \sum_{\nu, \mathbf{k},\sigma} V_{\nu}^{\phantom\dagger} \left({d}^{\dag}_{\nu \sigma} c^{\phantom{\dag}}_{\nu \mathbf{k}\sigma} + h.c.\right),
\end{equation}
with tunnel-coupling $V_{\nu}$ to lead-$\nu$. The strength of hybridization to lead-$\nu$ is embodied in
$\Gamma_{\nu} =\pi V^{2}_{\nu} \rho$, where $\rho = N/(2D)$ is the (uniform) total
density of states of the lead, and $N$ the number of
lead-orbitals ($N \rightarrow \infty$ in the continuum limit).
The effect of a magnetic field is readily encompassed by adding
$\hat{H}_B = -g \mu_B B \hat{S}_z$, with
$\hat{S}_z =\frac{1}{2}\sum_{\nu}(\hat{n}_{\nu\uparrow}-\hat{n}_{\nu\downarrow})$
the total $z$-component of spin on the dots,
$B$ the magnetic field strength, and $g$ the electron $g$-factor. We have taken  $g=2$
in comparing to experiment, but note that a slightly smaller value
would lead to even better agreement between experiment and theory in Fig.~3 of the paper.

In our calculations we use the parameters $U, ~U', ~t$ and $\Gamma_{\nu}$ determined experimentally,
and a temperature $T \sim 100$ mK. Experimentally, the dot level energies are controlled by two gate
voltages, $V_{gL}$ and $V_{gR}$. One might naively anticipate $\epsilon_{\nu} \propto -V_{g\nu}$ but, as is physically natural given the experimental set-up (Fig.~1(a)), there is some `cross talk' between the side gate voltages, and to obtain good agreement between experiment and theory it is necessary to include some dependence of
$\epsilon_{L}$ on $V_{gR}$ and of $\epsilon_{R}$ on $V_{gL}$. Specifically, we find
\begin{equation}
\epsilon_L \propto - V_{gL} - aV_{gR},~~~~ \epsilon_R \propto - aV_{gL} - V_{gR},
\end{equation}
with $a = 0.4$ giving the best agreement.


\section{NRG calculation of conductance}
We use the full density matrix (FDM) formulation~\cite{Spete:06,
Sweic:07} of the NRG~\cite{Swils:75,Skris:80,Skris2:80} (for a
recent review see Ref.~\onlinecite{Sbull:08}), keeping around 2000
states per iteration. To reduce discretization error, we average
over different lead discretizations and calculate the dot Green
functions via the self-energy method~\cite{Sbull:98}. This enables
us to work reliably with a relatively large discretization parameter
of $\Lambda =9$.

To calculate the zero-bias ($V_{sd} = 0$) conductance, $G_c$, of the
model at finite temperature, we use an exact Kubo-type
current-current correlation function approach, derived via
linear-response theory~\cite{Ssaka:97,Ssaka:2000}
\begin{equation}
\label{eq:lr}
G_c(V_{sd} = 0, T) = \frac{e^2}{2h}\left[~\bar{\sigma}_{LL} + \bar{\sigma}_{RR} - 2\bar{\sigma}_{LR}~\right],
\end{equation}
where (with $\mu ,\nu \in \{ L,R\}$)
\begin{equation}
\bar{\sigma}_{\mu \nu} = \frac{\pi}{\hbar^2} V_{\mu}V_{\nu} ~\underset{\omega\rightarrow 0}{\mathrm{lim}}
 \left[\mathrm{Im} \frac{C_{\mu \nu} (\omega)}{\omega}\right]~.
\end{equation}
The correlation functions $C_{\mu\nu}(\omega)$, which can be calculated directly via the FDM-NRG, are given by
\begin{equation}
C_{\mu\nu}(\omega) = \int_{-\infty}^{\infty} d\omega ~ e^{-i\omega t} ~ (-i)~\theta(t)~\big\langle~[\hat{j}_{\mu}(t),\hat{j}_{\nu}(0)]_+~\big\rangle
\end{equation}
where $[\hat{a},\hat{b}]_+$ denotes an anticommutator and
$\theta(t)$ is the unit step function. The $\hat{j}_{\mu}$, defined by
\begin{equation}
\label{eq:curr}
\hat{j}_{\mu} = \sum_{\sigma} ~\left(f^{\dag}_{\mu 0 \sigma} d_{\mu \sigma} - d^{\dag}_{\mu \sigma} f_{\mu 0 \sigma} \right)~,
\end{equation}
are current operators (modulo constants), with $f_{\mu 0 \sigma} =
\frac{1}{\sqrt{N}}\sum_{\textbf{k}}c_{\mu \textbf{k} \sigma}$ the
operator for the zero-orbital of the Wilson chain representation of
lead $\mu =L,R$~\cite{Swils:75,Skris:80,Skris2:80}. At $T=0$,  the
zero-bias conductance eqs.\ref{eq:lr}-\ref{eq:curr} takes the
simpler, physically intuitive form~\cite{Sgeor:99}
\begin{equation}
G_c(V_{sd} =0,T=0) = \frac{8e^2}{h}\Gamma_L \Gamma_R|G_{LR} (\omega = 0)|^2~.
\end{equation}
Here $G_{LR}(\omega)$ is the off-diagonal (or interdot) retarded Green function,
$G_{LR}(\omega) \overset{F.T.}{\longleftrightarrow} -i\theta(t)\langle [ d^{\phantom\dagger}_{L\sigma}(t), d^{\dagger}_{R\sigma}]_{+}^{\phantom\dagger}\rangle$, which may also be calculated via the FDM-NRG; and
$\omega =0$ denotes the zero-bias Fermi level. Results for the $T=0$ conductance obtained this way agree
very well (as they should) with those calculated using eqs.\ref{eq:lr}-\ref{eq:curr} as $T \rightarrow 0$.

 At finite bias nothing exact can be said and, in practice, approximations must inevitably be made. Here
we make the common approximation of bias-independent self-energies
(which as judged by previous comparison to
experiment~\cite{Sgalp:10,Sande:08,Sloga:09} appears quite
successful), and approximate the finite-bias conductance by
\begin{equation}
G_c(V_{sd}, T)= \frac{8e^2}{h}\Gamma_L \Gamma_R \int_{-\infty}^{\infty}d\omega ~\left[ -\lambda\frac{\partial f_{L}(\omega)}{\partial \omega} -(1 - \lambda)\frac{\partial f_{R}(\omega)}{\partial \omega} \right]\\ |G_{LR} (\omega, T)|^2
\end{equation}
where $f_{\nu}(\omega) =[e^{(\omega -\mu_{\nu})/k_{B}T} +1]^{-1}$ is
the Fermi function for lead $\nu =L,R$, with chemical potentials
$\mu_{R} =\lambda eV_{sd}$ and $\mu_{L} =-(1-\lambda ) eV_{sd}$. The
quantity $\lambda$ ($\in [0,1]$) is a measure of how the voltage
bias is partitioned between the leads; $\lambda = 0.5$ being
perfectly symmetric, and $\lambda = 0$ or $1$ meaning the bias is
applied entirely on one lead. We take $\lambda = 0.8$ throughout,
this value being found to give the best agreement with the
experimental data (in particular for the slopes of the Coulomb
blockade peaks, which are known to depend on
$\lambda$~\cite{Slogan:JCP, Swright:11}).



\begin{thebibliography}{10}
\bibitem{J}
In the (1,1) regime, $J\simeq \frac{4t^{2}}{U-U'} [1-(\epsilon_{L}-\epsilon_{R})^{2}/(U-U')^{2}]^{-1}$, and thus increases with detuning $(\epsilon_{L}-\epsilon_{R})$.
\bibitem{faraway}
In our case by contrast, states such as $(2,0)$ or $(0,2)$ are ${\cal{O}}(U)$ higher in energy than the
$(1,1)$ states close to the center of the $(1,1)$ charge regime, and thus irrelevant.
\bibitem{coupling}
In general, varying the barrier gate voltage affects both the tunnel coupling $t$ and the coupling
to the leads. For the charge configuration here the main effect is on the lead couplings.
\end{thebibliography}
\end{document}